# QUOTATION FOR THE VALUE ADDED ASSESSMENT DURING PRODUCT DEVELOPMENT AND PRODUCTION PROCESSES

Alain Bernard[(1)], Nicolas Perry[(1)], Jean-Charles Delplace[(2)], Serge Gabriel[(2)]
[(1)] IRCCyN UMR CNRS 6597, Ecole Centrale de Nantes, 1, rue de la Noë, BP 92101, 44321 – NANTES Cedex 3 (F)
[(2)] SMC Colombier Fontaine, 2 rue du Doubs 25260 - COLOMBIER FONTAINE (F)

Abstract: This communication is based on an original approach linking economical factors to technical and methodological ones. This work is applied to decision process for mix production. This approach is relevant for costing driving system. The main interesting point is that the quotation factors (linked to time indicators for each step of the industrial process) allow the complete evaluation and control of, on one hand, the global balance of the company for a six months period and, on the other hand, the reference values for each step of the process' cycle of the parts. This approach is based on a complete numerical traceability and control of the processes (design and manufacturing of the parts and tools, mass production). This is possible due to numerical models and to feedback loops for cost indicators analysis at design and production levels. Quotation is also the base for the design requirements and for the choice and the configuration of the production process. The reference values of the quotation generate the base reference parameters of the process steps and operations. The traceability of real values (real time consuming, real consumable) is mainly used for a statistic feedback to the quotation application. The industrial environment is a steel sand casting company with a wide mix product and the application concerns both design and manufacturing. The production system is fully automated and integrates different products at the same time.

## 1.     INTRODUCTION

SMC COLOMBIER FONTAINE is a company in the AFE METAL group, which uses a sand casting process to manufacture steel primary parts. To reduce the "time to market", primary part producers need to reduce the time and cost of the industrialisation process. These factors, in addition to the global goal of improving process performance levels, brought SMC to develop numerical technologies and traceability from quotation to part delivery [1].

Nowadays, these improvements are incorporated into company culture. The next step in reducing the time and cost of the production process is to introduce a complete methodology of use and experience feedback of these new models and methods.

To be able to generalise this approach, a CAD methodology imposes and thus becomes a step in the industrialisation process. The amount of improvements engendered by the numerical technologies largely justifies the time investment made to obtain a numerical definition of all the different elements in the sand casting process [2].

The objective of our approach is to optimise the product and its production process by generating a complete numerical reference, through the integration of quotation, CAD, simulation, new manufacturing technologies and effective production processes.

This approach is the base of the integration of the three decision levels of a company, strategic, tactic and operational (Figure 1). At the strategic level, the quotation elaboration is directly linked to the economical impact of each production. At the tactic level and due to a complete numerical integration, quotation indicators are used to configure the processes parameters based on the CAD definitions of parts and tools. At the operational level, the This approach is the base of the integration of the three decision levels of a company, strategic, tactic and operational. At the strategic level, the quotation elaboration is directly linked to the economical impact of each production. At the tactic level and due to a complete numerical integration, quotation indicators are used to configure the processes parameters based on the CAD definitions of parts and tools. At the operational level, the processes are monitored and a complete numerical traceability and control is used in order to provide feedback of the value chain costs from design and production to the quotation application.

This paper presents the key factors of the approach's efficiency.

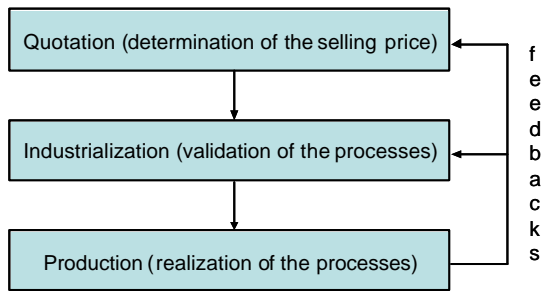

*Figure 1.* Main steps of the proposed approach

## 2. INDUSTRIAL CONTEXT OF THE PROJECT

SMC has a wide product range with 7 different automatic casting lines, about 1000 references per year and an average of 15 new tooling per month. Currently, during the sand casting process it takes from 20 to 40 hours to complete a CAD study; that is to say, model the part, the master pattern, the pattern-plates, the cluster and simulate the fill up and the solidification. At this time, generalizing this kind of study to all parts is possible due to the introduction of the new CAD methodology in the technical office. For this reason, toolmaker suppliers can use different moulding techniques to manufacture the pattern plates from the CAD models of the parts, but mainly from the CAD models of the tools. In this approach, the simulation's accuracy is hindered by the random repeatability associated with the tooling. All clusters are simulated and thus the results guide the designers to improve the rules supporting tooling definition. Because of the adequacy of the CAD model and the real cluster geometry, reliable simulation results are obtained. It optimises overall product industrialisation time and cost by reducing the iterations needed to perfect the serial production design. The use of a complete design methodology for both casting parts and tooling is one of the key factors of the project for the validation of a numerical engineering in sand casting. The performance factors give the fundamental orientations for the CAD design of parts and tooling and are completely process dependant. These factors are linked to the economical indicators that characterise the influence of a production on the economical balance of the company. The main difficulty is to have a dynamic interaction between strategic decision indicators and effective industrial processes behaviour (time, quality).

## 3. POSITION OF THE PROBLEM

The objective is to define on one hand, a complete numerical reference model of parts and tools and on the other hand, enterprise processes driven by costs estimations. In such a context, a design methodology for casting parts and tooling is indispensable in order to increase the efficiency of the design phase (decreasing time and cots) and to improve the quality control of the complete processes.

So, our approach refers to a proposition of a knowledge-based engineering approach [3][4][5] and also to a methodology for integrating economic criteria in design and production decisions [6].

A major problem of cost estimation comes from the increasing ratio of indirect cost in the global price of the part (mainly due to the evolutions of the market and the process cycle). Several analytics methods try to evaluate the exact price to be applied.

First, cost accounting gives no efficient solution to this problematic (mainly due to the lack of connections between the repartition keys of fixed costs and their real consumption).

Secondly, the ABC method (Activity Based Costing) focuses on cost drivers from the design to the delivery phase of the part cycle process [7][8][9]. Consequently, the cost estimation becomes more relevant, but this method is long to adjust and can be complex (each activity should be evaluate and associate with a cost driver).

Thirdly, the Value Added Unit method [10][11] has the same vision than ABC, but uses a single cost driver unit for all the steps. Its objective consists in determining the profitability of each sale. This becomes a very efficient tool to manage the effect of a part on the global economic balance of the enterprise. Our approach refers to this third method.

The cost definition is linked to intrinsic parameters of the product and time key-factors of enterprise processes. For example, the design phase is considered as generating uncompressible costs and is one of the heavier indirect charges for the product. So one of the main interests of our approach is to increase the efficiency of the technicians' work, that is to say, to increase the number of CAD definitions for a given period of time (increasing the productivity of the design phase).

Due to these considerations, the proposed approach can be considered as a valuable new approach for efficient cost-driven of enterprise processes. Quotation allows characterizing both a cost for a given batch of parts and the complete choices of the key factors of the industrial processes that will be used to produce this batch of parts.

The second very important point concerns the fact that all the processes are closed-loop monitored in order to readjust the quotation indicators depending on realistic time capabilities of the processes. Up to now, this updating data is manual but should be automated.

## 4. DEFINITION OF THE PROPOSED APPROACH

To be able to validate these concepts, SMC decided to act on different areas of the process' cycle of the casting products. This cycle is divided in three main steps: quotation, industrialization and production.

Quotation consists in evaluating the possibility to have benefit in realizing the production of a given part.

When this first step is validated, industrialization allows defining the raw (casting) part, the cluster and simulates the process parameters in order to validate the CAD definitions.

Then the production planning system takes into account the production order and defines the best repartition of the parts' flow on the automatic casting lines.

The main originality of our approach is that these three phases are completely connected and integrated to quotation indicators and the enterprise processes are driven coherently depending on these indicators.

### 4.1 QUOTATION CONTEXT AND OBJECTIVES

The industrial context is characterized by a large variety of parts. This mix-product has administrative consequences (large number of quotations, of commercial propositions, of client communications) and also industrial consequences (obligation to increase the design productivity, to improve the process efficiency, to decrease the scrap ratio, to monitor the real capability of the resources).

All these services and means have consequent effects on the costs. But, generally the market drives the price and the companies hardly integrate these costs for the parts. This point shows the importance of a very efficient costs management and the need of closed-loop interaction between quotation indicators' (that determine the processes parameters and characteristics) and the effective processes that are used for production.

Costs can be considered in two categories: fixed costs (that depends on salaries and enterprise structure) and variable costs (related to the number of manufactured parts, to the consumable used in the production processes like materials, energy, etc…).

According to the cost management method presented paragraph 3, each production can be considered as contributive (that covers both fixed and variable costs) or marginal (if it does not cover). The basic unit used for the computation is the time spent by activity. The economic objectives of the enterprise for the current year give an average's cost per hour for the global balance. With these data the quotation service can quickly decide to take or not a command, depending on the negotiation's results with the client and the corresponding price decided. The interest for the company is to have a maximum of contributive productions but some marginal ones are necessary in order to obtain the volume of production, a regulation of the mix product and the production scheduling (the company has to produce every day).

The main interest of our approach is to exploit the numerical reference model from this management strategic level. This level is based on three applicative layers, the quotation, the industrial and commercial planning and the dynamic stabilization point. The first level is necessary in order to evaluate the kind of production (contributive or not) and to measure the impact of such a production on the exploitation account. One should remind that the objective is not to determine the price of the production but to evaluate the contribution of the production by computing the corresponding hour's rate (which corresponds to the sell price of a production hour without consumables). This value can be compared to the ratio that has been determined as the base for a positive financial result of the company.

Of course, the quotation phase is based on a knowledge database that contains the major variables characterizing the enterprise processes with an Excel interface. This is a dynamic tool through the consequences immediate visualization of one of these variables modification (value of the hour rate). These functionalities allow mastering the economical characteristics of the manufacturing processes. Due to the complete integration with technical indicators of the processes, quotation step is possible in less than ten minutes and, when considering the past productions and when comparing the effective results with the previsions, the precision is less than 5%. Then the industrial and commercial plan indicates the contribution of the production to the enterprise result thanks to the comparison with the dynamic stabilization point.

The numerical reference model used for the quotation computation is also used for design and production stages. All the information is stored in the same knowledge database and is accessible for all the process cycle applications (from quotation to delivery).

## 4.2 PARTS AND TOOLING DESIGN METHODOLOGY

The second step of the numerical chain is the design of parts and tooling. This stage is based on a design methodology that allows validating the productions processes by taking into account the corresponding constraints.

At the beginning (Figure 2), the traditional manufacturing of a part was based on the definition of a 2D drawing and to manual fabrication of the tooling. Numerical integration that allows using the knowledge database has been developed and the technicians of the design office elaborate and validate each of the intermediately objects of design for parts and tooling. They also validate the production processes by using simulation software for temperature and metallurgical studies.

On a practical and methodological point of view, technicians use personal computers (laptops) which contain for each study simultaneously the quotation application and data, the knowledge database and the CAD application. Particular developments, based on the design methodology, have been implemented and each technician can validate coherently step-by-step the design process and the different models created for the parts and the tools.

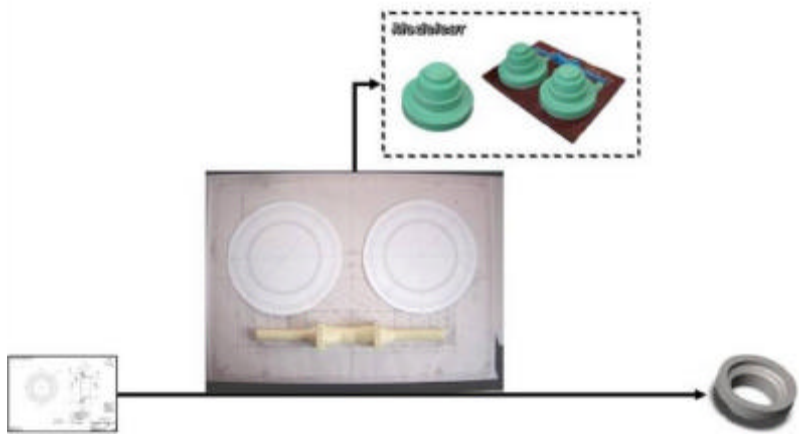

*Figure 2.* Conventional process

The design and simulation inputs and results are completely integrated with the knowledge database and are connected to the quotation factors. The definition of the processes corresponds to the decisions that have been validated during the quotation stage. The time of the design stage has been dramatically reduced (by four). The delivery time of the tooling (moulds, core boxes) is less than one week. As a result, cost has been decreased by a factor two.

All the numerical information needed for the production phase are automatically generated (tooling quality and topology, tooling manufacturing processes, production plan and parameters, casting parts and tooling references for traceability, numerical models of the cluster for the automatic cutting of the parts from the cluster, for the quality control of the tooling, for the quality control of the parts at the end of the production process). All the hypothesis concerning the production configuration are managed and closed-loop by the product planning application, based on the same knowledge database (Figure 3).

When the numerical simulation and validation of the process has been proceeded, all the technical and temporal data are sent to the production planning application in order to execute the production phase.

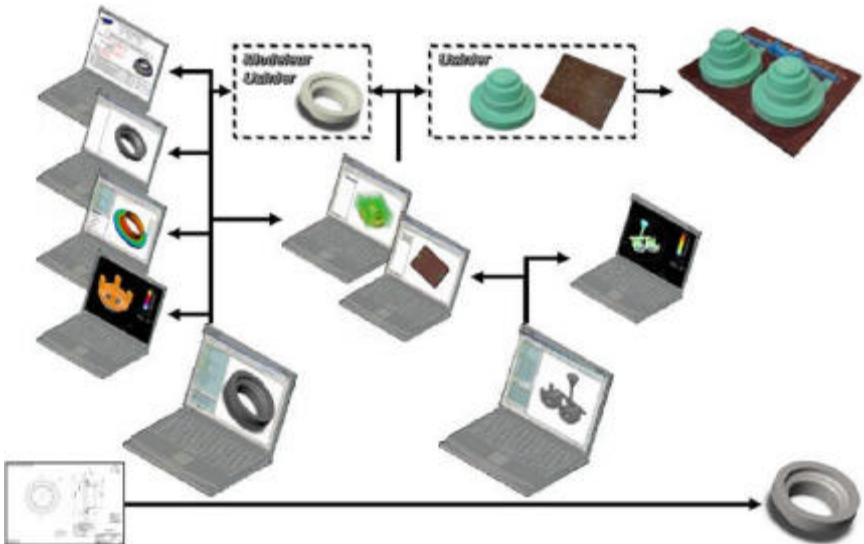

*Figure 3.* Actual numerical engineering process

## 4.3　　PRODUCTION NUMERICAL TRACEABILITY

The first phase of the manufacturing process is the manufacturing of few parts before the mass production phase. The benefit of such pre-production (good geometry, good material characteristics, good process) gives very important information about tooling geometry and capability, about the mass of the clusters and of the parts, about the cycle time, about the global robustness of the production process. This means that in priority the numerical referential model gives the insurance of a complete geometrical mastering.

This is possible due to the fact that each production sub process (mainly tooling geometry control and also core placement control) is closed loop. For example, the use of this unique referential model, based on what has been decided during quotation phase, allow controlling presence of cores and geometry and damages of joining surfaces. This last point is very important due to an important proportion of manual tasks dedicated to the suppression of exes of material (defaults on joining surfaces).

The decided process plan is traced using analogical codes. The operators validate the productions using a STL viewer that presents graphically the 3D graphical model of the parts. Then this is possible to evaluate the exact impacts of each production and let dynamically to re-actualize the economical factors of quotation indicators.

At present, SMC Colombier Fontaine company economical results are very positive due to this complete numerical integration from quotation to parts delivery. All the elementary processes (design and production) are completely closed loops and allow obtaining a global coherency for the mastering of the overall industrial activity at all company levels.

## 4.4 SYNTHESIS OF THE PROPOSED APPROACH

The following picture (figure 4) summarizes the complete process based on the quotation attributes and on the common, collective and distributive database that represents and memorizes the complete data related to each production. The time, from quotation to parts' delivery, represents in fact the main periods of the life-cycle of the casting parts. Sometime SMC asks subcontractors for part machining in order to deliver to the client finished parts, ready for assembly. During all these steps, data are memorized. These data are necessary and sufficient in order to be able to reproduce the same process in the future. These data, based on product models [12] for life-cycle [13], contribute to the increasing of the knowledge related to the company processes. These models are the key factors of the complete methodology proposed in this project.

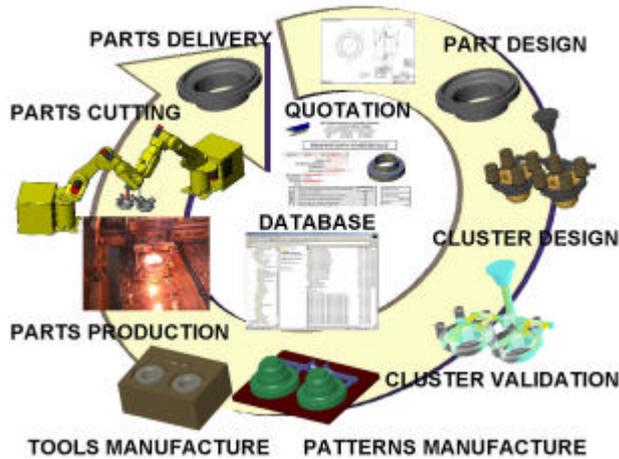

*Figure 4.* Complete industrial process based on two common references (quotation and collective database)

For each industrial study, the completion of the data is tested by the existence of wanted and effectively realized value chain parameters for all the industrial processes. The measure of the differences allows re-actualizing the quotation parameters. These differences are mainly related to quality and time. Concerning the design phase, time has been reduced due to the use of the complete methodology and to the structure of the database for all the parts and tools components (logical dependency between the part, the cores and the corresponding core-boxes, link between the part, the parting line and the corresponding geometry on the patterns and the tools, etc…).

## 5. CONCLUSIONS AND PERSPECTIVES

With the accomplishment of this study, SMC has introduced a complete economical and numerical integration. After the economical, numerical and technical validations, the payback highlights a very significant progress for costs, which confirms the successful introduction of knowledge-based elementary decision processes in the company. The design methodology proposed in this project reduces part design time and the CAD hourly rate. The design costs have been reduced enough thanks to the generalisation of the numerical models for tooling design. All of these improvements should facilitate SMC's generalised implementation of numerical integration and help them make their offer, in the field of sand casting, more competitive.